\documentclass[twocolumn,showpacs,superscriptaddress]{revtex4-2}
\usepackage{graphicx}
\usepackage{amssymb}
\usepackage{natbib}
\usepackage{calc}
\usepackage{textcomp}
\usepackage{bm}
\usepackage{color,soul}

\begin{document}

\title{Tracking the nematicity in cuprate superconductors: a resistivity study under uniaxial pressure}

\author{Tao Xie}
\affiliation{Beijing National Laboratory for Condensed Matter Physics, Institute of Physics, Chinese Academy of Sciences, Beijing 100190, China}
\affiliation{School of Physical Sciences, University of Chinese Academy of Sciences, Beijing 100190, China}
\affiliation{Neutron Scattering Division, Oak Ridge National Laboratory, Oak Ridge, Tennessee 37831, USA}
\author{Zhaoyu Liu}
\affiliation{Beijing National Laboratory for Condensed Matter Physics, Institute of Physics, Chinese Academy of Sciences, Beijing 100190, China}
\affiliation{School of Physical Sciences, University of Chinese Academy of Sciences, Beijing 100190, China}
\author{Yanhong Gu}
\affiliation{Beijing National Laboratory for Condensed Matter Physics, Institute of Physics, Chinese Academy of Sciences, Beijing 100190, China}
\affiliation{School of Physical Sciences, University of Chinese Academy of Sciences, Beijing 100190, China}
\author{Dongliang Gong}
\affiliation{Beijing National Laboratory for Condensed Matter Physics, Institute of Physics, Chinese Academy of Sciences, Beijing 100190, China}
\affiliation{School of Physical Sciences, University of Chinese Academy of Sciences, Beijing 100190, China}
\author{Huican Mao}
\affiliation{Beijing National Laboratory for Condensed Matter Physics, Institute of Physics, Chinese Academy of Sciences, Beijing 100190, China}
\affiliation{School of Physical Sciences, University of Chinese Academy of Sciences, Beijing 100190, China}
\author{Jing Liu}
\affiliation{Beijing National Laboratory for Condensed Matter Physics, Institute of Physics, Chinese Academy of Sciences, Beijing 100190, China}
\affiliation{School of Physical Sciences, University of Chinese Academy of Sciences, Beijing 100190, China}
\author{Cheng Hu}
\affiliation{Beijing National Laboratory for Condensed Matter Physics, Institute of Physics, Chinese Academy of Sciences, Beijing 100190, China}
\affiliation{School of Physical Sciences, University of Chinese Academy of Sciences, Beijing 100190, China}
\author{Xiaoyan Ma}
\affiliation{Beijing National Laboratory for Condensed Matter Physics, Institute of Physics, Chinese Academy of Sciences, Beijing 100190, China}
\affiliation{School of Physical Sciences, University of Chinese Academy of Sciences, Beijing 100190, China}
\author{Yuan Yao}
\affiliation{Beijing National Laboratory for Condensed Matter Physics, Institute of Physics, Chinese Academy of Sciences, Beijing 100190, China}
\author{Lin Zhao}
\affiliation{Beijing National Laboratory for Condensed Matter Physics, Institute of Physics, Chinese Academy of Sciences, Beijing 100190, China}
\author{Xingjiang Zhou}
\affiliation{Beijing National Laboratory for Condensed Matter Physics, Institute of Physics, Chinese Academy of Sciences, Beijing 100190, China}
\affiliation{School of Physical Sciences, University of Chinese Academy of Sciences, Beijing 100190, China}
\affiliation{Collaborative Innovation Center of Quantum Matter, Beijing 100190, China}
\affiliation{Songshan Lake Materials Laboratory, Dongguan, Guangdong 523808, China}
\author{John Schneeloch}
\affiliation{Condensed Matter Physics \& Materials Science Department, Brookhaven National Laboratory, Upton, NY 11973, USA}
\author{Genda Gu}
\affiliation{Condensed Matter Physics \& Materials Science Department, Brookhaven National Laboratory, Upton, NY 11973, USA}
\author{Sergey Danilkin}
\affiliation{Australian Centre for Neutron Scattering, Australian Nuclear Science and Technology Organization, Lucas Heights NSW-2234, Australia}
\author{Yi-feng Yang}
\affiliation{Beijing National Laboratory for Condensed Matter Physics, Institute of Physics, Chinese Academy of Sciences, Beijing 100190, China}
\affiliation{School of Physical Sciences, University of Chinese Academy of Sciences, Beijing 100190, China}
\affiliation{Collaborative Innovation Center of Quantum Matter, Beijing 100190, China}
\affiliation{Songshan Lake Materials Laboratory, Dongguan, Guangdong 523808, China}
\author{Huiqian Luo}
\affiliation{Beijing National Laboratory for Condensed Matter Physics, Institute of Physics, Chinese Academy of Sciences, Beijing 100190, China}
\affiliation{Songshan Lake Materials Laboratory, Dongguan, Guangdong 523808, China}
\author{Shiliang Li}
\email{slli@iphy.ac.cn}
\affiliation{Beijing National Laboratory for Condensed Matter Physics, Institute of Physics, Chinese Academy of Sciences, Beijing 100190, China}
\affiliation{School of Physical Sciences, University of Chinese Academy of Sciences, Beijing 100190, China}
\affiliation{Collaborative Innovation Center of Quantum Matter, Beijing 100190, China}
\affiliation{Songshan Lake Materials Laboratory, Dongguan, Guangdong 523808, China}

\begin{abstract}
Overshadowing the superconducting dome in hole-doped cuprates, the pseudogap state is still one of the mysteries that no consensus can be achieved. It has been suggested that the rotational symmetry is broken in this state and may result in a nematic phase transition, whose temperature seems to coincide with the onset temperature of the pseudogap state $T^*$ around optimal doping level, raising the question whether the pseudogap results from the establishment of the nematic order. Here we report results of resistivity measurements under uniaxial pressure on several hole-doped cuprates, where the normalized slope of the elastoresistivity $\zeta$ can be obtained as illustrated in iron-based superconductors. The temperature dependence of $\zeta$ along particular lattice axis exhibits kink feature at $T_{k}$ and shows Curie-Weiss-like behavior above it, which may suggest a spontaneous nematic transition. While $T_{k}$ seems to be the same as $T^*$ around the optimal doping and in the overdoped region, they become very different in underdoped La$_{2-x}$Sr$_{x}$CuO$_4$. Our results suggest that the nematic order, if indeed existing, is an electronic phase within the pseudogap state.	
\end{abstract}


\maketitle

\section{Introduction}

Many electronic orders such as electronic stripes and charge ordering \cite{TranquadaJM95,WuT11,GhiringhelliG12,NetoEH14,CominR14}, and nematic order that break the in-plane rotational symmetry from $C_4$ to $C_2$ \cite{AndoY02,HinkovV08,DaouR10,LawlerMJ13,AchkarAJ16,ChoiniereO15,ChoiniereO17,ZhengY17,WuJ17} have been observed in the pseudogap state in high-$T_c$ superconducting cuprates. Previous results from Nernst measurements show two types of nematicity in YBa$_2$Cu$_3$O$_{6+\delta}$ (YBCO) within the pseudogap state \cite{DaouR10,ChoiniereO15,ChoiniereO17}. The first type tracks the charge-density-wave (CDW) modulations around  hole concentration $p$ = 0.12 and the second one tracks the pseudogap energy with the onset temperature of nematicity $T_{nem}$ much lower than the onset temperature of the pseudogap state $T^*$ for $p <$ 0.11. However, torque-magnetometry measurements in YBCO ($p \geq$ 0.11) and HgBa$_2$CuO$_{4+\delta}$ (Hg-1201) provide thermodynamic evidence for the rotational symmetry breaking at $T^*$, suggesting the onset of pseudogap state is associated with a second-order nematic phase transition \cite{SatoY17,Murayama2019}. It is not clear whether these contradictory results come from the different techniques and standards in determining the relevant temperatures. What's more, Raman scattering measurements and elastoresistance measurements on Bi$_{2-y}$Pb$_{y}$Sr$_2$CaCu$_2$O$_{8+\delta}$ (Bi-2212) suggest the existence of a nematic quantum critical point around the endpoint of the pseudogap~\cite{Auvray2019,Ishida2020}. Resonant X-ray scattering study on the La$_{1.6-x}$Nd$_{0.4}$Sr$_x$CuO$_4$ indicates the vanishment of nematic order beyond the pseudogap state~\cite{Naman2021}. A recent temperature-dependent angle-resolved photoemission spectroscopy (ARPES) study of nematicity in slightly overdoped Bi-2212 shows that the nematicity is enhanced in the pseudogap state and is suppressed in the superconducting state~\cite{Nakata2021}. These studies highlight the correlation between the nematicity and pseudogap state, which means that nematicity could be an important part to understand the pseudogap and superconductivity in high-$T_c$ superconductors.

The studies on the nematic order in iron-based superconductors show that the spontaneous nematic transition can be well studied by measuring the elastoresistivity above the transition temperature~\cite{ChuJH12,KuoHH16,LiuZ16,GuY17}, which suggests that it may also provide key information in understanding the nematicity in cuprate superconductors.

Taking classical magnet as an example, the zero-field magnetic susceptibility should show a divergent behavior when approaching the transition temperature from the paramagnetic state. For the nematic transition, when the conjugated field is uniaxial pressure/strain \cite{FernandesRM14}, the nematic susceptibility can be obtained by measuring the uniaxial pressure/strain dependence of a physical property tracking the nematic order, such as resistivity. Indeed, elastroresistivity measurements on many iron-based superconductors show divergent behavior of nematic susceptibility above $T_{nem}$~\cite{ChuJH12,KuoHH16,LiuZ16,GuY17,Gong2017}, providing thermodynamical evidences for the nematic order. Compared to the direct measurement of resistivity anisotropy, the nematic susceptibility measurement has a much higher resolution and does not suffer from effect of residual strain from glue, etc. \cite{XiaoR15}, and the external pressure that is usually used to detwin the sample. Therefore, one may expect to observe a similar behavior of the nematic susceptibility in cuprates if there is a state that is indeed associated with a nematic phase transition \cite{NieL14}. A Curie-Weiss-like evolution of the nematic susceptibility was recently reported in Bi-2212 cuprates~\cite{Ishida2020}, and the potential divergent behavior of the nematic susceptibility around the $T^*$ suggests the presence of a nematic phase transition. Motivated by these existing elastoresistivity studies in iron-based superconductors and Bi-2212 cuprates, here we study the elastoresistivity in several kinds of cuprates to seek more experimental evidences that whether the pseudogap state in cuprates is correlated with a nematic transition or not.
\section{Experimental details}

In this study, we have chosen three classes of hole-doped cuprates, La$_{2-x}$Sr$_x$CuO$_4$ (LSCO), Bi$_{1.74}$Sr$_{1.88}$Pb$_{0.38}$CuO$_{6+\delta}$ (Bi-2201) and Bi$_{2-y}$Pb$_{y}$Sr$_2$CaCu$_2$O$_{8+\delta}$ ($y$ = 0 or 0.7), which were all grown by the traveling solvent floating zone method~\cite{Luo2007,Shen2009,Hucker2011,Liu2019}. The hole concentration $p$ is determined by the value of $T_c$ in Bi-2212 and Bi-2201 \cite{VishikIM12,KudoK09}, while that in LSCO is determined by the Sr doping level $x$. The orientations of the crystals were determined by X-ray Laue camera, single-crystal X-ray diffraction and scanning transmission electron microscope (STEM). The samples were cut into thin rectangular plates along either the Cu-O-Cu or diagonal direction [see Fig.~\ref{fig2}(a)] by a high-precision diamond wire saw. The uniaxial pressure was applied along the longer edge of the rectangle by a home-made device based on the piezo-bender as described in Ref.~\cite{LiuZ16}, which is able to avoid the effect of residual strain from glue and measure the resistivity change across zero pressure continuously by compressing and stretching the sample. The pressure applied on the sample is controlled by the voltage applied on the piezo bender~\cite{LiuZ16}. The standard resistivity and elastroresistivity were measured by the standard four-probe method on a Quantum Design Physical Property Measurement System (PPMS). The DC magnetic susceptibility measurements were performed on a Quantum Design Magnetic Property Measurement System (MPMS) with zero-field-cooling (ZFC) method. The neutron diffraction data of the $x$ = 0.17 LSCO sample were collected at the thermal neutron triple-axis spectrometer TAIPAN at Australian Centre for Neutron Scattering, ANSTO, Australia, with a fixed incident and final energy $E{_i}$ = $E{_f}$ = 14.87 meV.
\begin{figure}[tbp]
\includegraphics[width=\columnwidth]{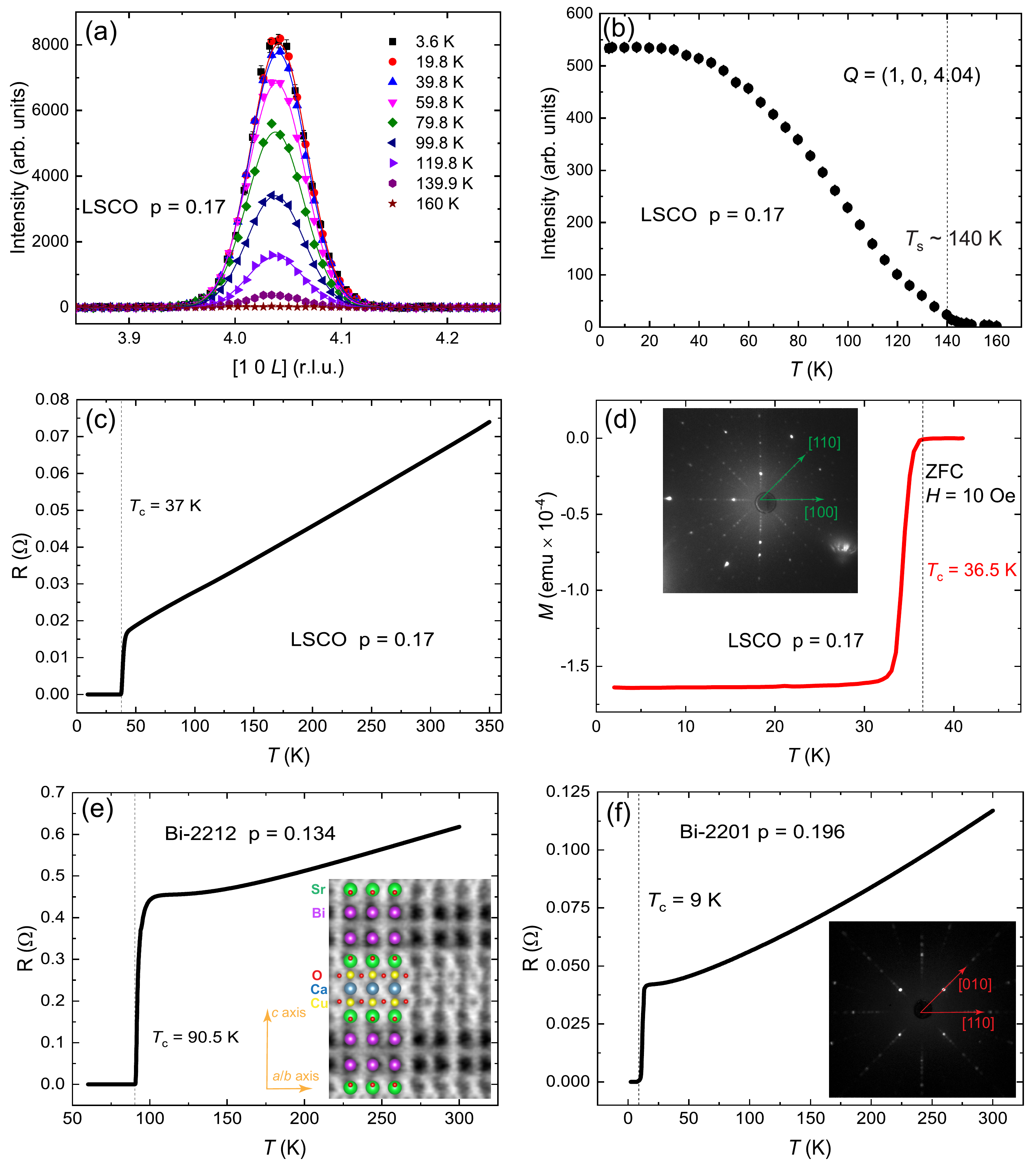}
\caption{Characterizations of our cuprate crystals. (a) Nuclear reflection (1 0 4) of the $x$ = 0.17 LSCO sample measured at a series of temperatures. (b) Temperature dependence of the peak intensity at $Q$ = (1 0 4). (c)--(d) Temperature dependence of resistance and magnetic susceptibility of the $x$ = 0.17 LSCO sample, which show sharp superconducting transition at $T_c$ $\sim$ 37 K. The inset of (d) is a typical Laue diffraction pattern of our LSCO crystal. (e)--(f) Two examples of the resistance-temperature curves of our Bi-2212 and Bi-2201 sample. The inset of (e) is a STEM ABF image of the Bi2212 crystal. The inset of (f) is a typical Laue diffraction pattern of the Bi-2201 crystal.}
\label{fig1}
\end{figure}

\section{Results and discussions}

The neutron diffraction measurements on a $x$ = 0.17 LSCO sample show that the nuclear Bragg peak at $Q$ = (1 0 4) (orthorhombic notation) loses its intensity with the increasing temperature and disappears completely above $\sim$ 140 K [Fig.~\ref{fig1}(a) and (b)], which indicates an orthorhombic-tetragonal structural transition at $T_s$ $\sim$ 140 K~\cite{Gilardi2002,FlemingRM87}. The resistance and magnetic susceptibility as a function of temperature show sharp superconducting transitions at $T_c$ $\sim$ 37 K for $x$ = 0.17 LSCO crystals [Figs.~\ref{fig1}(c) and \ref{fig1}(d)]. Two typical resistance-temperature curves of the Bi-2212 and the Bi-2201 sample are shown in Figs.~\ref{fig1}(e) and \ref{fig1}(f), respectively. The insets in Figs.~\ref{fig1}(d)--\ref{fig1}(f) present typical Laue diffraction patterns and STEM annular bright field (ABF) image of the LSCO, Bi-2212, and Bi-2201 crystals, respectively, which represent how the orientations of the crystals were determined. The high symmetry directions are indicated by arrows.

\begin{figure}[tbp]
\includegraphics[width=\columnwidth]{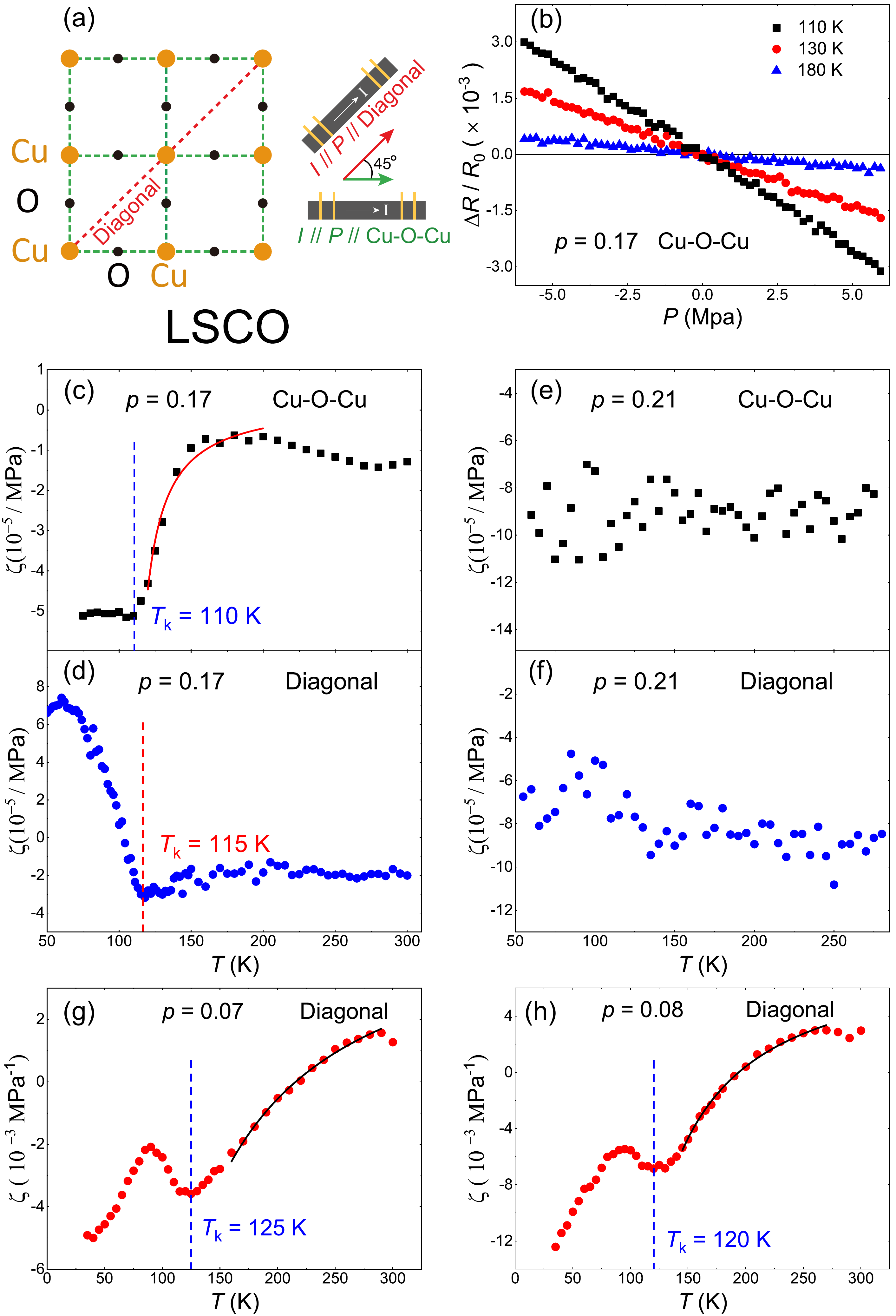}
\caption{Schematic of sample orientations and elastroresistivity results of the LSCO samples. (a) Left: The CuO$_2$ plane with the definition of the two directions. Right: Schematic of our samples with standard four probes cut along the two measured directions. $P$ and $I$ denote uniaxial pressure and electric current, respectively. (b) Selected pressure dependence of $\Delta R/R_0$ in the $x$ = 0.17 LSCO sample at several temperatures. (c)--(d) Temperature dependence of $\zeta$ in the $x$ = 0.17 sample for the uniaxial pressure along the Cu-O-Cu and diagonal directions, respectively. The red line in (c) is a Curie-Weiss-like fitting with $T'$ = 105.5 K. (e)--(f) Similar measurements on the $x$ = 0.21 sample. (g)--(h) Temperature dependence of $\zeta$ in the $x$ = 0.07 and 0.08 samples along the diagonal direction. The black lines in (e) and (f) are Curie-Weiss-like fittings with $T'$ = 31 K and 65.6 K, respectively. The dashed lines indicate the kink temperatures $T_k$. }
\label{fig2}
\end{figure}

The definition of the two used crystallographic directions: (i) Cu-O-Cu direction and (ii) diagonal direction in this study is illustrated in Fig.~\ref{fig2}(a). Figure~\ref{fig2}(b) shows the pressure (along the Cu-O-Cu direction) dependence of $\Delta R/R_0$ in the $x$ = 0.17 LSCO sample at several temperatures. The positive and negative pressures correspond to compressing and stretching the sample, respectively. Since the resistance shows linear pressure dependence, we define $\zeta$ as $d(\Delta R/R_0)/dP$, where $\Delta R$ and $R_0$ are the resistance change under pressure $P$ and the resistance at zero pressure, respectively. It has been shown that in iron-based superconductors the $\zeta$ can be defined as nematic susceptibility above the nematic transition if it is measured along the nematic ordering direction~\cite{LiuZ16,GuY17}, assuming that the resistivity change is mainly caused by nematic fluctuations.

Figure~\ref{fig2}(c) presents the temperature dependence of $\zeta$ along the Cu-O-Cu direction for the $x$ = 0.17 LSCO. The most promising features are the kink at $T_k$ = 110 K and the sharp increase of $|\zeta|$ above it. The solid line in Fig.~\ref{fig2}(c) is a Curie-Weiss-like fitting of the data as $\zeta$ = $A/(T-T')$+$y_0$, where $A$, $T'$ and $y_0$ are all temperature-independent parameters \cite{LiuZ16}. $T'$ is lower than $T_k$, which may be caused by the coupling between the electronic system and the lattice as suggested in iron-based superconductors \cite{ChuJH12,LiuZ16}. Below $T_k$, $\zeta$ becomes independent of temperature. The result along the diagonal direction [Fig.~\ref{fig2}(d)] also shows a kink at the similar temperature. Different from that along the Cu-O-Cu direction, $\zeta$ along the diagonal direction changes little above $T_k$ but dramatically below it. The small difference of $T_k$s for Cu-O-Cu and diagonal direction is most likely due to slightly inhomogeneous doping during the growth since these two samples have $T_c$ of 37 K and 36.5 K, respectively. It should be noted that the tetragonal-orthorhombic structural transition at about 140 K for this doping level [Fig.~\ref{fig1}(a)--(b)] seems to have no effect on the elastoresistivity data, suggesting that the resistivity difference between the orthorhombic axes can be neglected. Figures~\ref{fig2}(e) and \ref{fig2}(f) show the same analyses of $\zeta$ on the overdoped $x$ = 0.21 LSCO. No obvious temperature dependence of $\zeta$ can be seen along both directions and all the features in the $x$ = 0.17 sample disappear.

For the underdoped $x$ = 0.07 and 0.08 LSCO samples, similar kink feature and divergent behavior of $\zeta$ are also observed along the diagonal direction, as shown in Figs.~\ref{fig2}(g) and \ref{fig2}(h). At lower temperatures, we can find an additional kink feature, below which the $|\zeta|$ increases dramatically. The origin of this additional kink is currently unknown, while the influence from the superconductivity can be one of the candidates. The changes of $\zeta$ in these two samples are much larger than that in the $x$ = 0.17 LSCO. In a previous study of resistivity anisotropy in LSCO samples with lower doping levels ($x \leq$ 0.04), the resistivity along the orthorhombic $b$ direction is smaller than that along the orthorhombic $a$ direction at high temperature, i.e., $\rho_b < \rho_a$ \cite{AndoY02}, which suggests that the positive value of $\zeta$ at high temperature in the $x$ = 0.07 and 0.08 samples here may result from the domains change under uniaxial pressure. Interestingly, $\rho_b/\rho_a$ quickly increases with decreasing temperature and becomes larger than 1 at low temperature \cite{AndoY02}, which is also consistent with the sign change of $\zeta$ in our measurements.

\begin{figure}[tbp]
\includegraphics[width=\columnwidth]{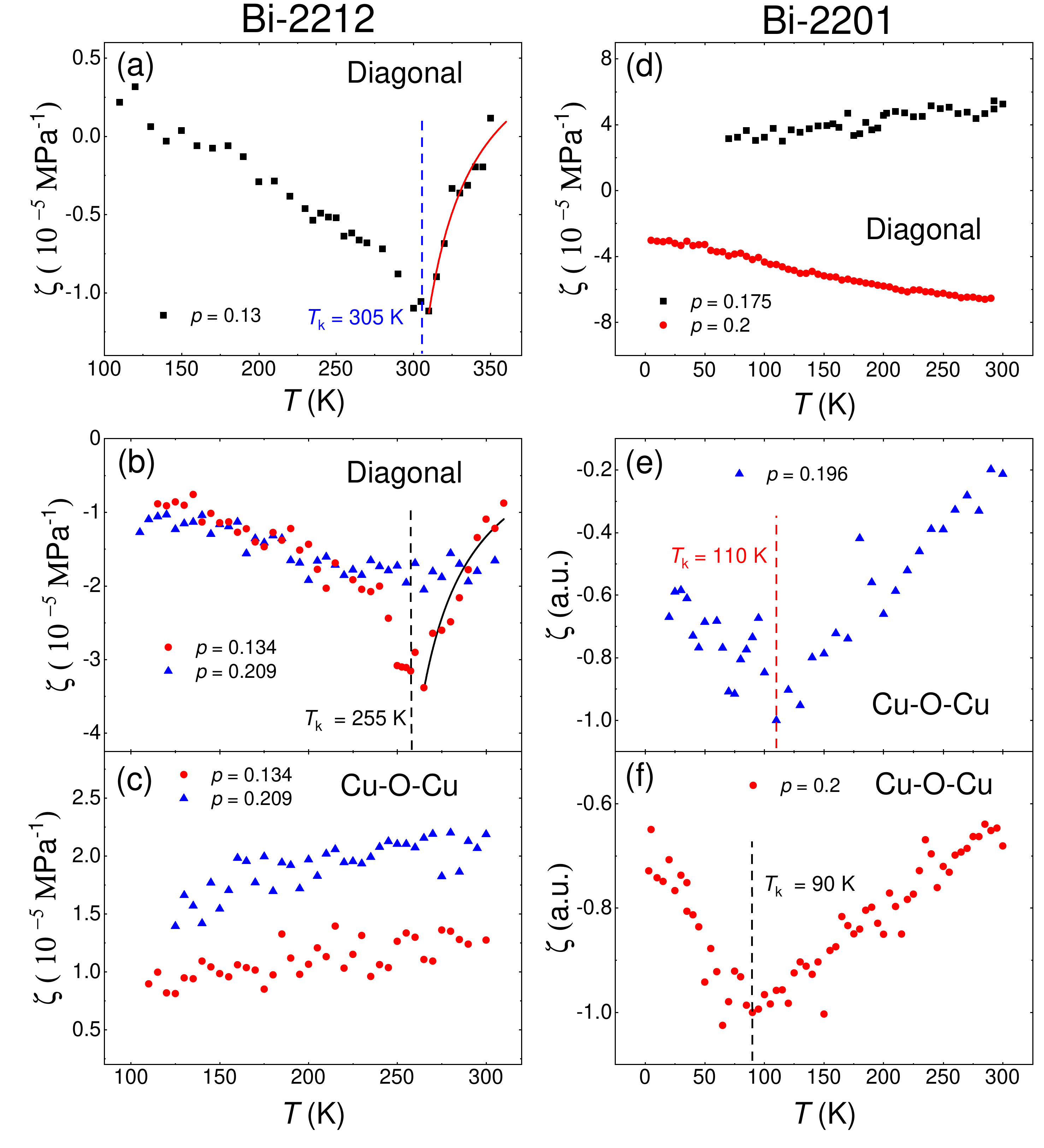}
\caption{Elastroresistivity results of the Bi-2212 and Bi-2201 samples. (a)--(c) Temperature dependence of $\zeta$ for the Bi-2212 samples along the diagonal direction and Cu-O-Cu direction. The solid lines in (a) and (b) are Curie-Weiss fittings with $T'$ = 278 K and 135 K, respectively. (d)--(f) Temperature dependence of $\zeta$ for the Bi-2201 samples along the diagonal direction and Cu-O-Cu direction. The dashed lines indicate the kink temperatures $T_k$.}
\label{fig3}
\end{figure}

Figures~\ref{fig3}(a)--\ref{fig3}(c) show results of the Bi-2212 samples, where again the kink features are found in the $p$ = 0.13 ($T_c$ $\sim$ 88.5 K) and $p$ = 0.134 sample ($T_c$ $\sim$ 90.5 K), and the direction showing the kinks is along the diagonal direction. Moreover, $|\zeta|$ increases rapidly with decreasing temperature above $T_k$, where the $\zeta$ can be fitted by Curie-Weiss-like function. $|\zeta|$ decreases slowly with decreasing temperature below $T_k$ [Figs.~\ref{fig3}(a) and \ref{fig3}(b)]. This behavior is reminiscent of the recent report on Bi-2212, where kinks and divergent behavior of nematic susceptibility were observed by measuring the elastoresistance under uniaxial strain~\cite{Ishida2020}. When the hole concentration level is increased to $p$ = 0.209 ($T_c$ $\sim$ 76.6 K), the kink feature disappears [Fig.~\ref{fig3}(b)]. For the $\zeta$ measurements along the Cu-O-Cu direction in both $p$ = 0.134 and 0.209 samples, no obvious temperature-dependent feature can be identified~[Fig.~\ref{fig3}(c)]. It should be noted that the crystallographic direction we observed the kink and divergence of $\zeta$ here (diagonal direction) is different from that (Cu-O-Cu direction) in Ref.~\cite{Ishida2020}.

In Bi-2201, we found the kink features of $\zeta$ along the Cu-O-Cu direction [Figs. \ref{fig3}(e) and \ref{fig3}(f)], which is just opposite to the case in the Bi-2212 samples [Figs. \ref{fig3}(a) and \ref{fig3}(b)]. However, above the kink temperature $T_k$, the increase of $|\zeta|$ with decreasing temperature is not as dramatic as those in LSCO and Bi-2212, so it cannot be fitted by the Curie-Weiss-like function. Interestingly, although no kink feature and divergent behavior of $\zeta$ can be observed along the diagonal direction, the $\zeta$ shows opposite sign in the $p$ = 0.175 and 0.2 Bi-2201 samples [Fig.~\ref{fig3}(d)].

These ubiquitous kink features of $\zeta$ along specific directions and its possible divergent behavior above $T_k$ suggest that nematic transition may be widely present in cuprates as that shown in iron-based superconductors \cite{ChuJH12,KuoHH16,LiuZ16,GuY17}, but very different behaviors of elastoresistivity are found among different materials. First, it seems that there is no one unified crystallographic direction to probe nematicity. The direction along which the kink feature and the divergent behavior (above $T_k$) of $\zeta$ can be observed is the Cu-O-Cu direction in LSCO ($x$ = 0.17) and Bi-2201, but it is the diagonal direction in Bi-2212 and very underdoped LSCO ($x$ = 0.07 and 0.08). This difference may be related to the crystal structure since both Bi-2212 and very underdoped LSCO are in the orthorhombic structure at room temperature \cite{TakahashiH94,IzquierdoM06}. It is consistent with the fact that the nematic direction may be affected by the crystal structure as shown in both iron-based \cite{FernandesRM14} and cuprate superconductors \cite{WuJ17}. It is worth noting that the nematicity in Bi-2212 observed from the scanning tunneling microscope (STM) measurements is along the Cu-O-Cu direction \cite{LawlerMJ13}, which is also different from our observations in Bi-2212 here and has been explained otherwise \cite{NetoE13}. Second, the elastoresistivity below $T_k$  behaves dramatically different with each other. The $|\zeta|$ along the nematic direction decreases slowly with decreasing temperature in Bi-2212 and Bi-2201, but is unchanged in the $x$ = 0.17 LSCO. In both $x$ = 0.07 and 0.08 LSCO, additional kink features can be observed below $T_k$. Third, the change of $|\zeta|$ with temperature in Bi-2201 is not as dramatic as those in LSCO and Bi-2212, probably due to the presences of very strong disorders in Bi-2201~\cite{KudoK09}. What's more, the change of $|\zeta|$ with temperature in all these studied cuprates are within one order of magnitude, which is a kind of small when compare with that in iron-based superconductors, where the change of $|\zeta|$ can cross two orders of magnitude~\cite{LiuZ16,GuY17,ChuJH12,KuoHH16}.

\begin{figure}[tbp]
\includegraphics[width=\columnwidth]{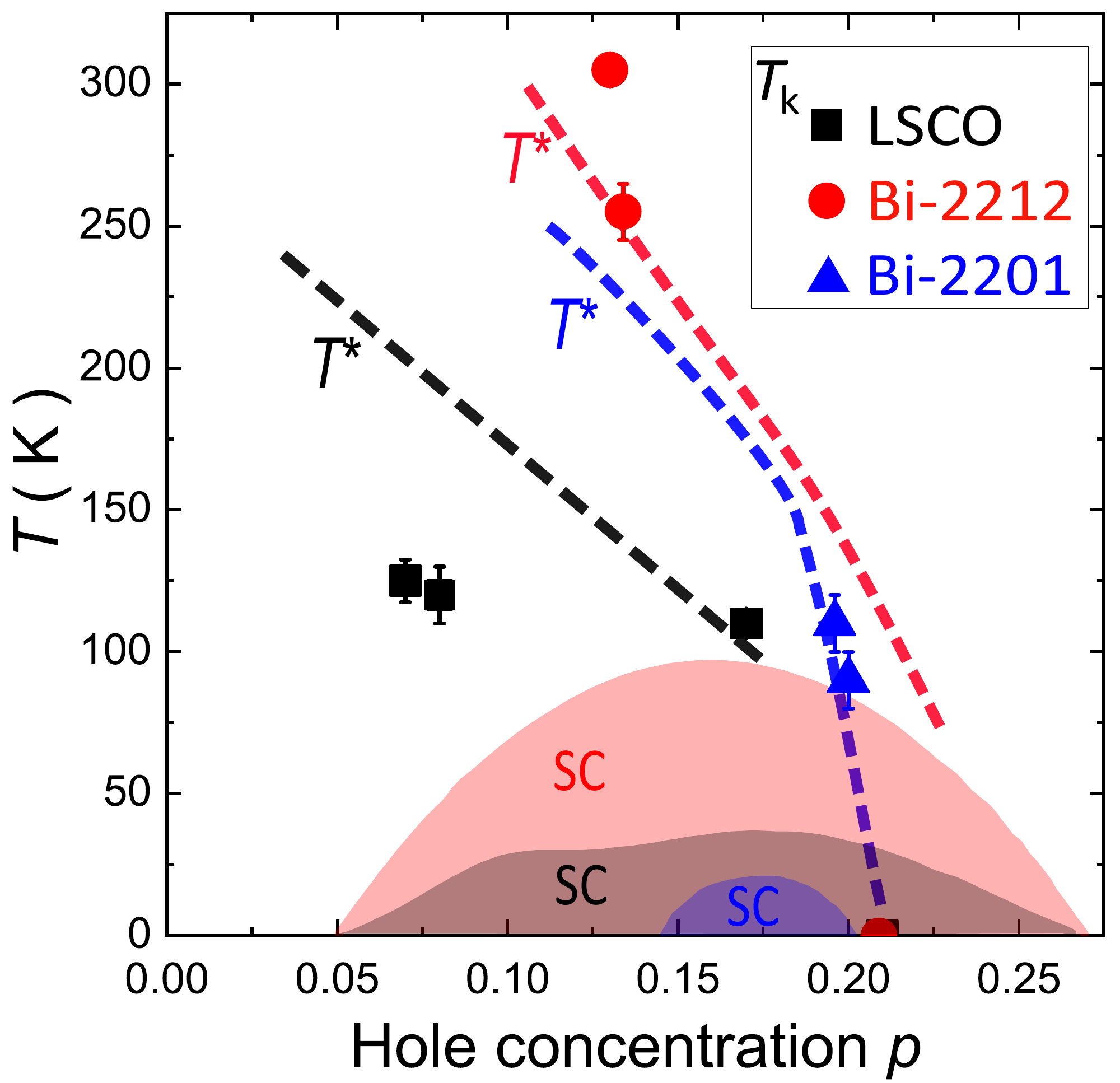}
\caption{Schematic phase diagram of cuprates. The black, red and blue dashed lines (shadow areas) represent the upper bounds of $T^*$ (superconducting domes) for LSCO, Bi-2212, and Bi-2201 \cite{ChoiniereO17,VishikIM12,KudoK09}, respectively. The solid square, circle, and triangle symbols are the kink temperature $T_k$s of $\zeta$ for LSCO, Bi-2212 and Bi-2201, respectively. The values of $T_k$ for the $x$ = 0.21 LSCO and $p$ = 0.209 Bi-2212 are set to zero. The vertical error bars on the symbols are estimated uncertainties of the corresponding $T_k$. }
\label{fig4}
\end{figure}

Despite the above differences among these cuprate supercondoctors, the kink temperature $T_k$ seem to be related to the onset temperature of pseudogap $T^*$, especially around the optimal and overdoped region. Figure~\ref{fig4} gives the schematic phase diagram of these three classes of cuprates~\cite{ChoiniereO17,VishikIM12,KudoK09}, where the $T_k$s are almost the same with $T^*$s around the optimal doping level and in the overdoped region. As described above, in the samples of LSCO and Bi-2212 with kink features, the $\zeta$ above $T_k$ can be fitted by the Curie-Weiss-like function to some extent, suggesting a divergent behavior like that observed in iron-based superconductors \cite{ChuJH12,LiuZ16}. It seems to be consistent with the suggestion that the onset of the pseudogap state in hole-doped cuprates is associated with a spontaneous nematic transition \cite{SatoY17}. However, this picture is not valid if we consider the underdoped LSCO ($x$ = 0.07 and 0.08), where $T_k$s are much lower than $T^*$s. Moreover, $\zeta$ shows no feature around $T^*$ ($\sim$ 200 K) [Figs.~\ref{fig2}(g) and \ref{fig2}(h)]. From this point of view, if the $T_ks$ correspond to nematic transitions, the nematic order should be a state within the pseudogap state, especially in the very underdoped region. This is consistent with the previous Nernst measurements on very underdoped YBCO~\cite{ChoiniereO15} and the resonant X-ray scattering study on underdoped La$_{1.6-x}$Nd$_{0.4}$Sr$_x$CuO$_4$~\cite{Naman2021}, where the nematic temperature has been shown to be significantly lower than $T^*$. This proposal seems to be also supported by a recent study of the anisotropy of resistivity, Seebeck coefficient, and Peltier coefficient in an underdoped YBCO~\cite{nonematicity2022}, where the nematicity is suggested to be absent at the $T^*$ and the low-temperature (well below the $T^*$) Peltier anisotropy is considered as nematicity and attributed to the development of CDW order.

\section{Summary}

Our elastoresistivity studies in several classes of cuprate superconductors widely observe kink features of $\zeta$ along particular crystallographic directions. The divergent behavior of $\zeta$ above $T_k$s may point to nematic transitions. While the nematic transition may happen at $T^*$ around optimal doping level and overdoped region, it becomes significantly lower than $T^*$ for the very underdoped cases. Therefore, the nematic order in cuprates may be just another phase within the pseudogap state, such as the stripes and CDW order~\cite{TranquadaJM95,WuT11,GhiringhelliG12,NetoEH14,CominR14,nonematicity2022}, which makes it as one of the competing or intertwined orders \cite{KeimerB15,FradkinE15}. Compared to the CDW state, the nematic phase can exist at much higher temperature and lower doping at least in LSCO and YBCO, which suggests a very close relationship between these two orders \cite{FradkinE15,KivelsonSA98}. Overall, our studies provide a wide perspective on the nematicity and pseudogap in cuprates, although some questions like the non-uniform nematic directions still need to be answered by further studies.

\section*{Acknowledgements}

This work is supported by the Ministry of Science and Technology of China (Grants Nos. 2020YFA0406003, 2021YFA1400401, 2018YFA0704200, 2017YFA0303100, 2017YFA0302903, 2015CB921302, and 2015CB921303), the National Natural Science Foundation of China (Grants Nos. 11674406, 11774401, 11522435, 11822411, and 11961160699), the Strategic Priority Research Program(B) of the Chinese Academy of Sciences (Grants Nos. XDB33010100, XDB25000000, and XDB07020000),  and the K. C. Wong Education Foundation (Grants No. GJTD-2020-01). Work at Oak Ridge National Laboratory (ORNL) was supported by the U.S. Department of Energy (DOE), Office of Science, Basic Energy Sciences, Materials Science
and Engineering Division. Work at Brookhaven was supported by the Office of Basic Energy Sciences (BES), Division of Materials Sciences and Engineering, U.S. Department of Energy (DOE), through Contract No. DE-SC0012704. J. S. is supported by the Center for Emergent Superconductivity, an Energy Frontier Research Center funded by the U.S. Department of Energy (DOE), Office of Science. H. L. and Y. Y. are supported by the Youth Innovation Promotion Association of CAS.

%

\end{document}